\documentclass[]{aa}  

\usepackage{natbib}
\bibpunct{(}{)}{;}{a}{}{,}
\usepackage{graphicx}
\usepackage{revsymb}	
\usepackage{amsmath}	
\usepackage[usenames]{color}	
\usepackage{epstopdf}	
\usepackage{txfonts}
\usepackage{amssymb}
\usepackage{color}
\usepackage[justification=centering]{caption}
\usepackage{geometry} 
\usepackage[parfill]{parskip} 
\usepackage{amssymb}
\usepackage{slantsc}
\usepackage{array}
\usepackage{url}
\usepackage{amsfonts}
\usepackage[colorlinks=true,linkcolor=blue, urlcolor=blue, citecolor=blue]{hyperref}
\usepackage{sidecap}
\usepackage{graphicx}
\usepackage{xcolor}
\usepackage{nicefrac}
\usepackage{subfigure}
\usepackage{epsfig}
\colorlet{rouge}{red!70!darkgray}

\begin{document}
\titlerunning{The age of the Methuselah star}
\title{The age of the Methuselah star in light of stellar evolution models with tailored abundances}
\author{C.~Guillaume\inst{1}\and G.~Buldgen\inst{1}\and A.~M.~Amarsi\inst{2}\and M.~A.~Dupret\inst{1}\and M.~S.~Lundkvist\inst{3}\and
J.~R.~Larsen\inst{3}  \and R.~Scuflaire\inst{1} \and A.~Noels\inst{1}}
\institute{STAR Institute, Université de Liège, Liège, Belgium \and Theoretical Astrophysics, Department of Physics and Astronomy, Uppsala University, Box 516, 751 20 Uppsala, Sweden \and Stellar Astrophysics Centre, Department of Physics and Astronomy, Aarhus University, 8000 Aarhus C, Denmark}
\date{June, 2024}
\abstract{HD140283, or the Methuselah star, is a well-known reference object in stellar evolution. Its peculiar chemical composition, proximity and absence of reddening makes it an interesting case-study of Pop II stars. Thanks to recent observational efforts, we now have precise interferometric and spectroscopic constraints, as well as revised astrometric parallaxes from the Gaia mission.} 
{We aim at determining the age of HD140283 with these lastest constraints, as well as quantifying the impact of systematics from physical inaccuracies in the stellar evolution models.}
{Using recent spectroscopic abundances from the literature, including 3D non-LTE values for C, O, and Fe,  we compute opacity tables specific to HD140283. We then use them in grids of stellar evolution models coupled to a Markov Chain Monte Carlo tool to determine the age of HD140283.}
{With our tailored models we find an age of 12.3Gy.  Using a solar-scaled mixture instead results in an age value of 14Gy, in tension with the age of the universe ($13.77\pm0.06$Gy). We also find that reducing the mixing length parameter from its solar calibrated value will lead to an even lower age, in  agreement with other recent studies. However, we find no direct evidence to favour a lower mixing length parameter value from our modelling.}
{Taking into account the specific elemental abundances is crucial for the modelling of HD140283, as it leads to significant differences in the inferred age. However, this effect is degenerate with a lowering of the mixing length parameter. In this respect, asteroseismic constraints might play a key role in accurately deriving the mass of HD140283, therefore strongly constraining its age.} 
\maketitle
\section{Introduction}

Population II stars, characterized by their low metallicity, are among the oldest objects in the Galaxy. They are typically found in the Galactic halo, globular clusters, and central regions of galaxies.  They are key fossils for studying the chemical evolution of our Galaxy and for Galactic Archaeology, including for indirect studies of primordial stars, the so-called Population III \citep[e.g.][]{Frebel2015}. The ages of these stars is of fundamental importance in this endeavour \citep[e.g.][]{Xiang2022}, but determining reliable stellar ages is  a major challenge as it depends on various physical ingredients in stellar evolution models that are often highly uncertain \citep{Lebreton2014I, Lebreton2014II,Ying2023}.

In this context the G-type subgiant HD140283, also called the Methuselah star, is an ideal benchmark old, metal-poor star due to its brightness, \textbf{absence} of reddening, proximity, and well-determined surface chemical composition. Numerous studies have carried out stellar evolution modelling of this star, leading to some debate in the community about its age. For example, \citet{Bond2013} and \citet{VandenBerg2014} find ages of $14.46 \pm 0.31 \ \text{Gyr}$ and $14.27 \pm 0.38 \ \text{Gyr}$, respectively, in tension with the age of the  Universe of $13.77 \pm 0.06 \ \text{Gyr}$ as determined from precise measurements of the Cosmic Microwave Background (CMB) and accurate estimations of the Hubble constant and baryon acoustic oscillations \citep{Planck,PlanckErr}. More recent work have alleviated this tension: \citet{Creevey2015} find $13.7\pm0.7$Gy, while \citet{Joyce2018} find a range between $12.5$ and $14.9$Gy.

In a more recent study, \citet{Tang2021} further revised the age of HD140283 down to $12.5\pm0.5$Gy, making use of the precise fundamental parameters recently presented by \citet{Karovicova2020}. In their study, they stressed the sensitivity of their models on the mixing length parameter, and arrive at a calibrated value that is $10$-$20\%$ below the solar value. However, this study does not discuss in detail the role of the stellar composition on the age inference. According to \citet{Bond2013} and \citet{VandenBerg2014}, this has a significant impact on the stellar evolution modelling, in particular via the O abundance.

To rectify this, we decided to compute tailored models of HD140283, investigating again the importance  of its chemical composition in light of  several spectroscopic  abundance analyses recently presented in the literature, in combination with the precise fundamental parameters reported  by \citet{Karovicova2020}. We also study the impact of non-standard physics, including turbulent diffusion, electronic screening, radiative opacities and non-solar mixing length parameter, on the inferred age of HD140283.

We start by presenting our observational constraints in Sect. \ref{Sec:Obs}, the grids of models and the physics they include in Sect. \ref{Sec:Models} and discuss the impact of various systematics in Sect. \ref{Sec:Discussion}. 

\section{Observational constraints}\label{Sec:Obs}

We present the fundamental parameters  in Table \ref{taObsConstraints}.  Most of the constraints come from \citet{Karovicova2020} who provided the radius, effective temperature, surface gravity and luminosity from  interferometric data (via the PAVO instrument at the CHARA array) combined with photometry, spectroscopy, isochrone fitting and Gaia astrometry. These constraints are fully consistent with \citet{Amarsi2019}.

We took individual elemental abundances from literature spectroscopic studies. The absolute abundance of Fe was taken from \citet{Amarsi2022}, which employs three-dimensional (3D) radiation-hydrodynamical models of stellar atmospheres, together with spectrum synthesis with departures from local thermodynamic equilibrium (non-LTE) taken into account. Similarly, the 3D non-LTE abundances of O (the most important element for the modelling of HD140283) and C were taken from \citet{Amarsi2019}: for these elements, the reported $\mathrm{[X/H]}$  values were adopted (together with their uncertainties), and converted to the absolute scale using  the solar composition of \citet{Asplund2021}, hereafter AAG21, adding the uncertainties in quadrature.

Many of the remaining elements were taken from the high-resolution study of \citet{Siqueira2015}. This study adopts 1D hydrostatic model atmospheres, but with departures from LTE taken into account for some elements: Na, Mg, Al, K, Ca, Sr, and Ba. These abundances were taken from the values of $\mathrm{[X/Fe]_{adopted}}$ reported in their Table 6, with an assumed uncertainty of $0.1$ dex. These were converted into an absolute scale using the Fe abundance from \citet{Amarsi2022} and the AAG21 solar composition, adding the uncertainties in quadrature. The 1D LTE abundance of S was taken from the value of $\mathrm{[S/Fe]}$ reported in \citet{Nissen2007}, with an uncertainty of $0.05$ dex from the line-by-line standard error, and converted in a similar way. The abundances of all remaining elements were calculated by scaling the AAG21 solar chemical composition using the Fe abundance from \citet{Amarsi2022}, with an enhancement of $\mathrm{[\alpha/Fe]}=0.4$ for Ne and Ar; the abundances of these remaining elements were assumed to have an uncertainty of $0.2$ dex.

\begin{figure}
	\centering
		\includegraphics[width=8.5cm]{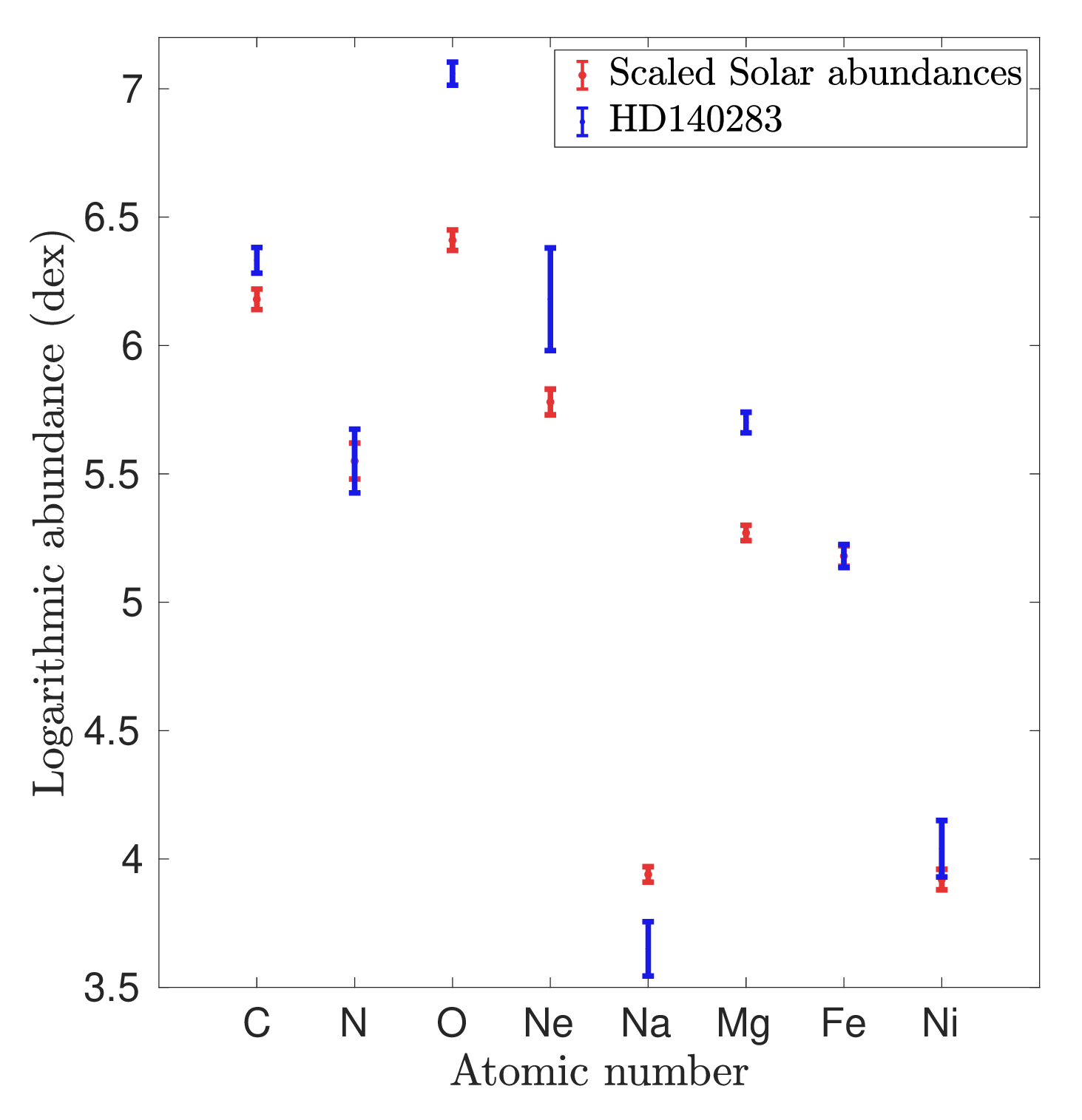}
    \caption{Abundances of the main 
    contributors to metal mass and to opacity for HD140283,
    based on literature spectroscopic values,
    as well as on a solar-scaled compilation
    (with no enhancement for $\alpha$-elements).}
	\label{Fig:Abundances}
\end{figure}

Table \ref{taObsConstraints} shows the metal mass fraction for the star $\left[\rm{M}/\rm{H}\right]=-1.82$, based on the specific chemical composition of HD140283, and that is used in our tailored models. In what follows, when using solar-scaled abundances from  AAG21 in our modelling, we will consider $\left[\rm{Fe}/\rm{H}\right]=\left[\rm{M}/\rm{H}\right]=-2.29\pm0.14$, taken from \citet{Karovicova2020}, which is close to the value of $\left[\rm{Fe}/\rm{H}\right]=-2.28$ advocated in \citet{Amarsi2022}. The solar-scaled composition is determined by adding $\left[\rm{M}/\rm{H}\right]=-2.29\pm0.14$ to the absolute abundances of AAG21, without any $\alpha$-enhancement. We illustrate in Fig. \ref{Fig:Abundances} the abundances we used alongside the solar abundances on the usual logarithmic scale,  $\log(\epsilon_{\mathrm{X}})=12+\log(N_{\mathrm{X}}/N_{\mathrm{H}})$,  with $N_{\mathrm{X}}$ and $N_{\mathrm{H}}$ the respective number densities of element X and hydrogen.

\begin{table}[h]
\caption{Observational constraints used in the modelling$^{1,2}$}
\label{taObsConstraints}
  \centering
\begin{tabular}{r | c }
\hline \hline
\textbf{Name} & \textbf{Value}\\ \hline
L (L$_{\odot}$) & $4.731 \pm 0.178$ \\
$\left[\rm{M}/\rm{H}\right]$ (dex) & $-1.82\pm0.07$\\ 
$\log$ g (dex) & $3.653\pm0.024$ \\
T$_{\rm{eff}}$ (K) & $5792\pm55$ \\
R (R$_{\odot}$) & $2.167\pm0.041$ \\
\hline
\end{tabular}
\end{table}

\section{Modelling and results}\label{Sec:Models}

\begin{figure*}
	\centering
		\includegraphics[width=16cm]{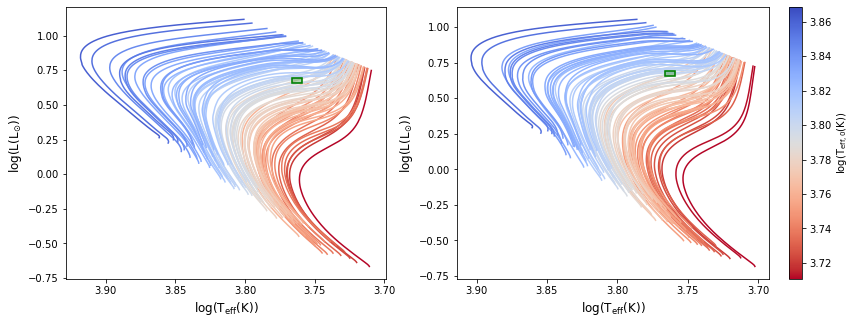}
	\caption{Left panel: HR diagram of our grid for the solar-scaled mixture
    with a solar-calibrated mixing-length parameter. Right panel: HR diagram of
    our grid for the tailored mixture of HD140283 with a solar-calibrated mixing-length parameter. The observational constraints for HD140283 are indicated by the green box.}
		\label{Fig:Grids}
\end{figure*}

We computed six separate grids of stellar evolution models using the Li\`ege Stellar Evolution code (CLES, \citealt{ScuflaireCles}), that we coupled with the Stellar Parameters INferred Systematically (SPInS) Markov Chain Monte Carlo modelling software \citep{Lebreton2020}. The grids span a mass range between $0.66$M$_{\odot}$ and $1.02$M$_{\odot}$, with a step of $0.01$, a metal mass fraction, Z, range from $5\times10^{-5}$ to $0.0018$ with a step of $2\times10^{-4}$. The initial hydrogen abundance is determined by assuming that the initial helium abundance is the primordial value, fixed to $0.251$ as in \citet{VandenBerg2014}. For all grids, we applied a cut at $3$R$_{\odot}$ once the model has exhausted the hydrogen in its core, based on the interferometric radius measurement available, and removed the pre-main sequence evolution has these can be excluded for HD140283. We can note in Fig. \ref{Fig:Grids} that this does not mean that some of the lower mass solutions are not already climbing the red giant branch, meaning that the geometry of the grid is quite complex and shows the potential for degeneracies in the inferred solution.

The first set of three grids was computed using a solar scaled mixture from \citet{Asplund2021}, together with the OPAL opacities \citep{OPAL} supplemented at low temperature by the Ferguson opacities \citep{Ferguson}, taking into account both the effects of miscroscopic diffusion \citep{Thoul,Paquette} and turbulence at the base of the convective envelope, considering a Vernazza Model C solar atmosphere \citep{Vernazza}. The difference between the three grids is the value of the mixing length parameter, $\alpha_{\rm{MLT}}$. One grid used a solar-calibrated mixing length value, using a classical solar calibration method \citep[see][]{Bahcall1982} using a standard solar model and three parameters (initial X, Z and $\alpha_{\rm{MLT}}$) calibrated against three constraints ($\rm{L}_{\odot}$, $\left(Z/X\right)_{\odot}$ and $\rm{R}_{\odot}$) with the same input physics as the grid. The other two grids applied a correction from the analytical formulae based on averaged 3D simulations from \citet{Magic2015}: this correction was based on either the entropy of the adiabat, or the entropy jump between the model and the averaged 3D simulation; they lead to lower value by $6.5\%$ and $9\%$ from the solar calibrated value, respectively.

The second set of three grids was computed using the  tailored abundances of HD140283 (see Sect. \ref{Sec:Obs}). We computed dedicated OP \citep{Badnell2005} and low Temperature AESOPUS opacities \citep{Marigo2009,Marigo2022} for this specific composition. We also considered for this grid the effects of microscopic diffusion and turbulence at the base of the convective zone. As before, the only parameter varied between the grid is the mixing-length parameter, again using a solar-calibrated value and two different corrected values via \citet{Magic2015} as before. The solar calibration in this case was performed using the AAG21 abundances in the calibration and the same input physics as that of the grid and the same constraints and free parameters as mentioned above. The use of custom abundances and opacities leads to a higher metallicity than in the solar-scaled case but also affects the details of the radiative transfer in the stellar interior (number of available electrons, ion populations, involved transitions, ...) that will affect the evolution of the star\citep[see][for detailed illustrations of the contributors to opacity in the solar case]{Blancard2012}.

Turbulent diffusion has been noted by \citet{VandenBerg2014} to be necessary to reproduce the final surface abundances of HD140283. A preliminary analysis using CLES models agrees with this, with microscopic diffusion being dominated by gravitational settling and quickly leading to a too high depletion of metals at the surface. While we neglected radiative accelerations in our work, it seems that they are unlikely to be sufficient to prevent the settling of metals in this specific case,  given that \citet{VandenBerg2014} included them in their work using the detailed treatment of the Montreal-Montpellier code. Therefore, we included the effects of turbulent mixing using a simplified
coefficient of \citet{Proffitt1991} 
\begin{equation} 
    D_{T}(r)=D \left(\frac{\rho_{\rm{BCZ}}}{\rho}\right)^{n}, 
\label{Eq:Turb} 
\end{equation}
with $D$ and $n$ being free parameters to be calibrated. We kept values close to solar ones, with $D=1200$ and $n=1.3$ \citep{Eggenberger2022}. {In practice, the need for additional mixing at the base of the convective envelopes has been discussed in various contexts \citep[see amongst others][]{Talon2006,Castro2007,Vick2013,Nordlander2024}}. In practice, such mixing could be calibrated from the observed lithium depletion. However, given the uncertainty in the initial value for HD140283 between the primordial lithium abundance \citep{Pitrou2018} and that of the Spite plateau \citep{Spite1982}, such a calibration would not be univocal. We come back on the impact of turbulent diffusion in Sect. \ref{Sec:Discussion} when testing various turbulence efficiencies.

We illustrate in Fig. \ref{Fig:Grids} our solar-scaled and dedicated abundance grids on the Hertzsprung-Russell diagram, assuming the solar-calibrated mixing length value with the corresponding physics. The box shows the position of HD140283, and as expected, switching from the solar scaled metallicity to the actual surface abundances, the tracks are significantly shifted, due to the higher metallicity resulting from the high oxygen enrichment leading to a higher metallicity, also directly impact our the radiative opacity in the stellar interior. Thus, we can already foresee a change in the inferred solution.

The parameters inferred by SPInS for each grid are provided in Table \ref{tabResulsFinal}. We used L, $\log$ g, R and $\left[\rm{M}/\rm{H}\right]$ as constraints for the fit, 10 walkers on 10 temperatures with 3000 burn-in steps and 10000 iterations in the MCMC procedure. 

\begin{table*}[h]
\caption{Optimal parameters inferred from SPInS}
\label{tabResulsFinal}
  \centering
\resizebox{\linewidth}{!}{
\begin{tabular}{r | c | c | c | c | c | c | c | c | c | c}
\hline \hline

\textbf{Name}&Mass (M$_{\odot}$)& Age (Gyr)&$\rm{X}_{0}$&$\rm{Z}_{0}$& Radius (R$_{\odot}$)&L (L$_{\odot}$)&T$_{\rm{eff}}$ (K)&log g (dex)&$\left[ \rm{M}/\rm{H} \right]$ (dex)\\ \hline
 AAG21, $\alpha_{\rm{MLT}\odot}$& $0.751\pm0.015$ &$14.10\pm0.98$ &  $0.749$ & $1.09\times10^{-4}$& $2.158$ &$4.72$&$5806$ &$3.648$&$-2.29$\\
AAG21, $\alpha_{\rm{MLT},\odot}-6.5\%$&  $0.757\pm0.015$ & $13.78\pm0.98$ & $0.749$& $1.09\times10^{-4}$ & $2.158$ &$4.72$&$5792$&$3.648$&$-2.29$\\ 
AAG21, $\alpha_{\rm{MLT},\odot}-9\%$&$0.760\pm0.015$ &$13.60\pm0.98$& $0.749$& $1.13\times10^{-4}$&$2.160$ & $4.73$& $5790$ &$3.649$&$-2.29$\\
HD140283, $\alpha_{\rm{MLT},\odot}$&$0.772\pm0.015$&$13.08\pm0.85$&  $0.749$& $3.11\times10^{-4}$&$2.170$&$4.70$&$5766$&$3.648$&$-1.82$\\
HD140283, $\alpha_{\rm{MLT}}-6.5\%$&$0.778\pm0.015$&$12.73\pm0.91$&$0.749$&$3.20\times10^{-4}$&$2.174$&$4.69$&$5760$&$3.654$&$-1.82$\\
HD140283, $\alpha_{\rm{MLT},\odot}-9\%$&$0.780\pm0.015$&$12.60\pm0.88$&$0.749$&$3.32\times10^{-4}$  &$2.177$ &$4.68$&$5752$&$3.654$&$-1.82$\\
\hline
\end{tabular}}
\end{table*}
 
When comparing our results to \citet{Bond2013} and \citet{VandenBerg2014}, we note that they found an age of $14.46 \pm 0.31 \ \text{Gyr}$ and $14.27 \pm 0.38\ \text{Gyr}$ in conflict with the age of the Universe, while taking into account the non-solar oxygen abundance of HD140283. Our solution differs from theirs in this respect, as we find a more massive solution and a lower age. Thiscan be explained from the difference in the parallax used. Here, as in \citet{Tang2021}, we use the Gaia parallax (via \citealt{Karovicova2020}). The Gaia DR3 parallax is $\pi = 16.26\pm 0.026$ mas, much lower than both the Hipparcos value and Hubble Fine Guidance Sensor of $\pi = 17.16\pm0.68$ mas and $\pi = 17.18\pm0.26$ mas. Thus, a smaller parallax implies that HD140283 is closer and would thus be more massive, thus younger, than previously thought. Indeed, they found a mass of $0.75\rm{M}_{\odot}$ while we find a more massive solution closer to that of \citet{Tang2021} (who found $0.81\pm0.05\rm{M}_{\odot}$) at $0.79\rm{M}_{\odot}$.

Compared to \citet{Tang2021}, our results provide a similar solution for the inferred mass and age.  However, we do not find evidence of favouring a lower than solar value for the mixing-length parameter. Indeed, all our computed grids could reproduce the observational constraints, whatever the $\alpha_{\rm{MLT}}$ value and display unimodal posterior distributions (see Sect. \ref{sec:Appendix}).  Reducing the value of $\alpha_{\rm{MLT}}$ leads to a more massive solution and thus a younger age for HD140283.  Indeed, for our solar-scaled solution, we can already see that further reducing this parameter could bring us in agreement with the results of \citet{Tang2021}. Owing to the degeneracies with chemical composition,  we actually recover their solution with our tailored abundance grid and solar-scaled mixing length parameter, because reducing $\alpha_{\rm{MLT}}$  has a similar effect to increasing the metallicity for low-mass stars. 

The degeneracy between mixing length parameter and chemical composition underlines the need for additional constraints for this star to determine whether departures from solar-calibrated values of $\alpha_{\rm{MLT}}$ are actually needed. In that respect, asteroseismic constraints might play a key role as they would offer strong constraints on the mass of the star, thus the duration of its core H-burning phase.

\section{Systematics affecting the inferred age}\label{Sec:Discussion}

In their paper, \citet{Tang2021} have investigated the impact of convective overshooting on the inferred age for HD140283 and found that it had no impact on their final results. This is in line with the results of \citet{Deheuvels2010} and \citet{Buldgen2019} who found that the effects of out of equilibrium burning of $^{3}$He did not impact the final inferred age in their studies. However, as noted by \citet{Tang2021}, additional effects such as diffusion in the radiative zone might actually lead to significant differences in the inferred age. 

\begin{figure}
	\centering
		\includegraphics[width=7.5cm]{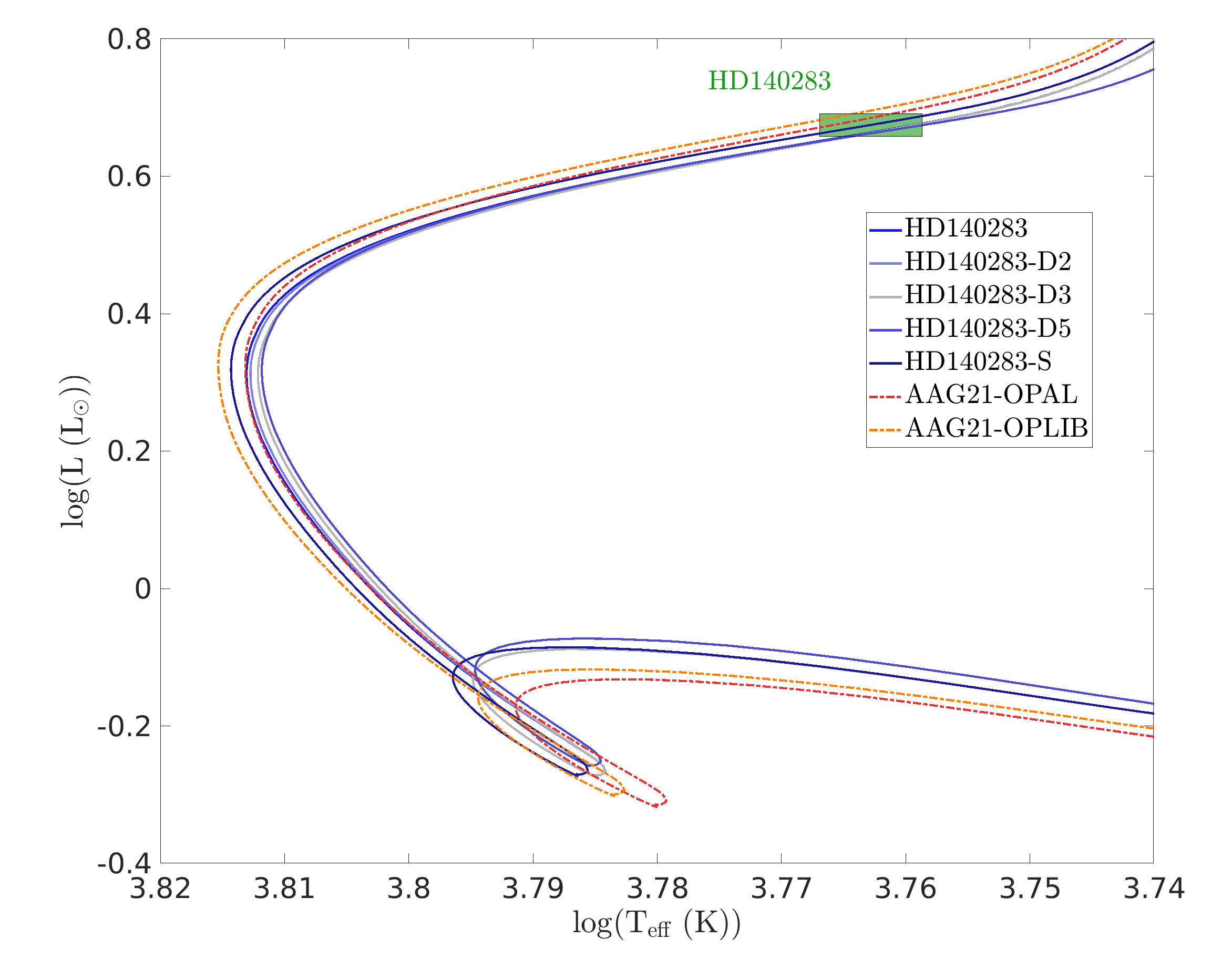}
	\caption{HR diagram of the models including variations of physical ingredients (diffusion, electronic screening, opacities) for HD140283.}
		\label{Fig:TrackSys}
\end{figure}

Here, we supplemented their study by investigating the effect of turbulent diffusion, varying the opacity table and considering the effects of dynamical screening of the electron gas in the core. Both effects lead to various degrees of changes in the thermodynamical conditions in the core that affects the evolutionary tracks and the duration of the main-sequence of HD140283. We plot the various evolutionary tracks for all our test cases in Fig. \ref{Fig:TrackSys}. To test if such variations had a significant impact on our solution, we used the parameters provided by SPInS and recomputed the evolutionary sequence varying each physical ingredient at a time and seeing whether the new model still allowed to reproduce all observational constraint within $1\sigma$ at a given age. 

The effects of microscopic diffusion are tested by varying the coefficient, D, and exponent, n, of Eq. \ref{Eq:Turb}. In Fig. \ref{Fig:TrackSys}, these models are shown as 'HD140283-Dn' with n the various exponent value we chose and the reference , noted $'HD140283'$, one being $n=1.3$ following prescriptions aiming at reproducing the behaviour of stellar models including angular momentum transport by the magnetic Tayler instability. As mentioned above, various efficiencies have been inferred in the past, by varying the simple parametric formulation used here, we can assess the impact of the efficiency of turbulence on the inferred age. Almost all tracks go through the box of the observational constraints in the HR diagram. We find that the surface chemical composition inferred from spectroscopy and the interferometric radius can be reproduced simultaneously. Therefore, the solution found by SPInS is not strongly affected by the variations we considered. We find very small changes in the age of the model, mostly due to a similar efficiency of the turbulent transport coefficients despite the variations of the exponent in Eq.\ref{Eq:Turb}, such an effect as also been seen for solar models  \citep[e.g.][for a recent discussion]{Buldgen2024}.  

To test the impact of the opacities, models were calculated using OPLIB opacities \citep{Colgan} instead of OPAL for the solar scaled solution\footnote{OPLIB tables for the mixture of HD140283 were not available.} which use a different framework to compute the interaction between high-energy radiation and the stellar plasma \citep[see][for a detailed comparison in solar models]{Buldgen2019Sun}.  Similar to what was found in the previous test for microscopic diffusion, this model reproduces the constraints within $1\sigma$ without the need to change its initial mass or chemical composition. The age found using the OPLIB opacities is of $13.70$Gy,$\approx 0.4$Gy younger than the one found using OPAL, meaning that a change in radiative opacities in the bulk of the radiative region (here a lowering due to the intrinsically lower opacity in high temperature of OPLIB compared to OP and OPAL) might be a relevant candidate to reduce the age of HD140283. 

Finally, we tested the impact of the electronic screening of the nuclear reactions. We simply switched off the screening by assuming, as suggested by \citet{Mussack2011A} and \citet{Mussack2011B}, that the overall effect would be similar. As with the two previous tests, this model reproduces all observational constraints within $1\sigma$ without the need to change the initial conditons, due to the main effect being a slight change in the central temperature at the ignition of hydrogen burning due to the lowering of the screening by electrons. The change in optimal age found is minimal, namely less than $0.05$Gy.

Our results thus show that none of these three effects significantly alter the conclusions of our study and that the main discrepancy in age using our modelling framework is obtained by varying the mass of the star. As discussed in Sect. \ref{Sec:Models}, such variations can be obtained by varying the mixing-length parameter, but we find no evidence to prefer a given value in our grid-based modelling. Future modelling using recent descriptions of convection might also provide interesting insights on this question \citep[e.g.]{Jorgensen2019, Manchon2024}, given the key role of HD140283 as a potential testbed for deviations from solar calibrated values of convective efficiency. It is also likely that allowing the initial helium mass fraction to vary within the existing primordial values would also lead to small changes in the inferred mass and therefore the age of HD140283.

\section{Conclusion}

In this study, we have carried out a new inference of the age of the so-called ``oldest star in the Universe'', HD140283, a.k.a.~the Methuselah star using individual spectroscopic abundances and custom opacity tables at high and low temperatures taking into account the high oxygen enrichment of the star. We used the SPInS MCMC modelling software coupled with grids of Li\`{e}ge stellar evolution models to study the impact of using a non-solar calibrated mixing length parameter value, using fixed analytical corrections from \citet{Magic2015}, alongside both a solar-scaled mixture and a tailored mixture from literature spectroscopic analyses. 

We have shown that taking into account the specific elemental abundances changes the inferred age from $\approx14$Gy to $\approx13$Gy. This effect of the composition is degenerate with the effect of the mixing-length parameter: models with a tailored composition and a solar-calibrated value of $\alpha_{\rm{MLT}}$ give rise to a similar age for HD140283 as models with a solar-scaled composition and a reduced $\alpha_{\rm{MLT}}$, such as those used by \citet{Tang2021}. Compared to \citet{Bond2013} and \citet{VandenBerg2014} that included the high oxygen abundance of the star but found an age in conflict with that of the Universe, we find that our younger solution results from the lower Gaia parallax (also used in \citet{Tang2021}) compared to both Hubble and Hipparcos values used in \citet{Bond2013} and \citet{VandenBerg2014}. 

We find no clear evidence of conflict with the age of the Universe when using the Gaia astrometric solutions, CHARA interferometric radius and  tailored spectroscopic abundances We also investigated the impact of systematics in  the models (turbulence, electronic screening and opacities) and found that they lead to  only small variations in the inferred age. Nevertheless, we consider that asteroseismic observations would be required to accurately constrain the mass of HD140283 and thus its age, potentially hinting at a departure from a solar-calibrated mixing length parameter value. Such constraints would allow to to a detailed analysis like that of \citet{Huber2024} for further test our stellar models on some of the oldest objects of the Milky Way. Regarding seismic constraints, we predict significantly different mean density values (thus different large frequency separations) and frequency of maximum power between the solar-scaled and oxygen-enriched solution, namely $0.1049 \rm{g/cm^{3}}$ and $0.1077\rm{g/cm{3}}$ and $495.68$ $\mu$Hz and $510.84$ $\mu$Hz. The availability of asteroseismic data may also allow for detailed comparisons using various evolution codes, as was done both in hare $\&$ hounds exercises\citep{Reese2016,Cunha2021} and detailed asteroseismic studies of \textit{Kepler} targets\citep{Silva2017}.

\begin{acknowledgements}
G.B. is funded by the Fonds National de la Recherche Scientifique (FNRS). A.M.A. gratefully acknowledges support from the Swedish Research Council (VR 2020-03940) and from the Crafoord Foundation via the Royal Swedish Academy of Sciences (CR 2024-0015). This work was supported by a research grant (42101) from VILLUM FONDEN as well as The Independent Research Fund Denmark’s Inge Lehmann program (grant agreement No. 1131-00014B).
\end{acknowledgements}

\bibliography{biblioarticleHD140283}
\begin{appendix}

\section{Posterior distributions for the MCMC runs}\label{sec:Appendix}
\begin{figure}
	\centering
		\includegraphics[width=9cm]{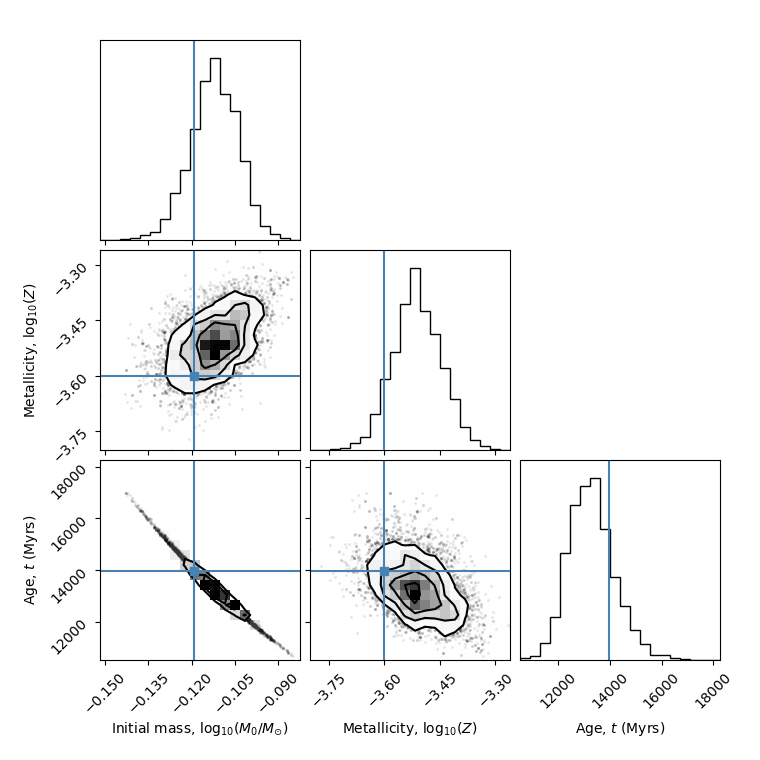}
	\caption{Posterior distributions from the SPInS runs for the logarithm of the mass, the age and the initial metal mass fraction  using a solar-calibrated mixing length parameter and the detailed element abundances.}
		\label{Fig:TriangleHD140283}
\end{figure}

\begin{figure}
	\centering
		\includegraphics[width=9cm]{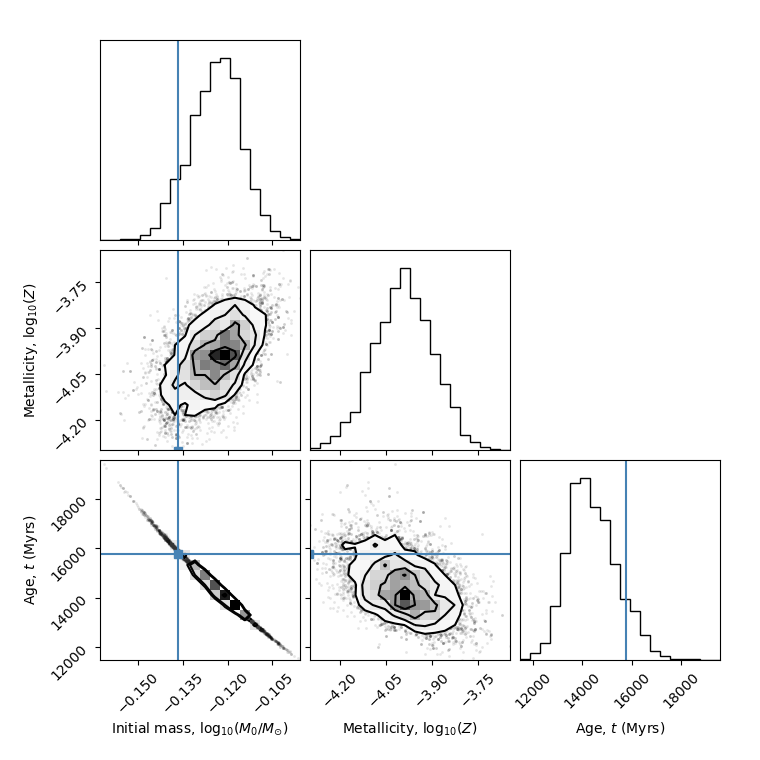}
	\caption{Posterior distributions from the SPInS runs for the logarithm of the mass, the age and the initial metal mass fraction  using a solar-calibrated mixing length parameter and the a solar-scaled mixture.}
		\label{Fig:TriangleScunScaled}
\end{figure}

Posterior distributions for the SPInS runs using both the solar scaled abundances, in Fig. \ref{Fig:TriangleScunScaled} and Fig. \ref{Fig:TriangleHD140283} the detailed spectroscopic abundances of HD140283 for a solar-scaled mixing-length parameter. 

\end{appendix}

\end{document}